\documentclass[useAMS,usenatbib]{mn2e}

\usepackage{amsmath}
\usepackage{amssymb}
\usepackage{multicol}
\usepackage{rotating}
\usepackage{lscape}

\usepackage{subfigure}
\usepackage{graphicx}
\usepackage{color}
\usepackage{placeins}
\usepackage{dblfloatfix}

\newcommand{\beq}{\begin{equation}}
\newcommand{\eeq}{\end{equation}}
\newcommand{\beqa}{\begin{eqnarray}}
\newcommand{\eeqa}{\end{eqnarray}}

\definecolor{gray}{gray}{0.55}

\begin{document}

%
\title[SDSS-III LOWZ Mocks]{
The clustering of galaxies in the SDSS-III Baryon Oscillation Spectroscopic Survey: mock galaxy catalogues for the low-redshift sample}
\author[Manera et al. ]{\parbox{\textwidth}{Marc Manera$^{1,2}$\thanks{email:marc.manera@port.ac.uk},
Lado Samushia$^{2,3}$, 
Rita Tojeiro$^{2}$,
Cullan Howlett$^{2}$,
Ashley J. Ross$^{2}$, 
Will J. Percival$^{2}$, 
Hector Gil-Mar\'in$^{2}$,
Joel R. Brownstein$^{4}$,
Angela Burden$^{2}$,
Francesco Montesano$^{5}$.
} 
\vspace*{3pt}\\
$^{1}$University College London, Gower Street, London WC1E 6BT, UK\\
$^{2}$Institute of Cosmology and Gravitation, Portsmouth University, Dennis Sciama Building, PO1 3FX, Portsmouth, UK\\
$^{3}$National Abastumani Astrophysical Observatory, Ilia State University, 2A Kazbegi Ave., GE-1060 Tbilisi, Georgia\\
$^{4}$Department of Physics and Astronomy, University of Utah, 115 S 1400 E, Salt Lake City, UT 84112, USA\\
$^{5}$Max-Planck-Insitut f\"{u}r Extraterrestrische Physik, Giessenbachstraße, 85748 Garching, Germany\\
}

\maketitle
%
\begin{abstract}
%
We present one thousand mock galaxy catalogues 
for the analysis of the Low Redshift Sample (LOWZ, effective redshift $z\sim 0.32$) 
of the Baryon Oscillation Spectroscopic Survey Data Releases 10 and 11.  
These mocks have been created following the \mbox{PTHalos} 
method of \cite{Manera13} revised to include new developments. The main improvement 
is the introduction of a redshift dependence in the 
Halo Occupation Distribution in order to account for the change of the 
galaxy number density with redshift. These mock catalogues are used in  
the analyses of the LOWZ galaxy clustering by the BOSS collaboration.  
\end{abstract}


\section{Introduction}

The Baryon Oscillation Spectroscopic Survey \cite[BOSS,][]{Dawson} is an spectroscopic 
survey that uses imaging data from SDSS-III \citep{Eis11} to map over 1.35 million galaxies 
covering an unprecedented volume of the universe over an area of aproximately a quarter
of the sky. The BOSS Data Release 11 \cite[DR11,][]{Aardwolf2013} contains $1,277,503$ galaxies 
covering $8 498$ square degrees, which, assuming a concordance $\Lambda\textrm{CDM}$ model, 
results in an effective volume of $8.4$ Gpc$^3$, the largest ever surveyed at this
density. 

BOSS targets two distinct galaxy samples: the CMASS sample, 
a high redshift sample $0.4 \lesssim z \lesssim 0.7 $ that selects galaxies
with roughly a constant stellar mass, and a low redshift sample $0.2 \lesssim z \lesssim 0.45 $ that targets
galaxies following an algorithm close to that designed for Luminous Red Galaxies (LRG) in SDSSS-I/II. 
Each of the DR11 samples has been used to fit the position of the baryon acoustic oscillation (BAO) feature,
constraining the distance measurement $D_V$ at two per cent for LOWZ \citep{Tojeiro2013} 
and one per cent in CMASS \citep{Aardwolf2013}. 
The latter is the most precise distance constraint ever obtained from a galaxy survey.    

The generation of mock galaxy catalogues is an essential component to the analysis of the data 
from any galaxy surveys.
Mocks are required for an 
accurate understanding of the sampling errors and the systematic errors of the clustering measurements,
including the effects of cosmic variance, non-linear evolution, scale-dependend bias, 
redshift distortions, and discreteness effects. They also enable detailed testing of analysis pipelines.
For a particular survey, mock galaxy catalogues
mimic the survey geometry, the number density of objects and their selection. 
All key science analyses of large scale galaxy clustering from BOSS DR9 relied heavily 
on the mock catalogues presented in \cite{Manera13}
Science analyses of ongoing and future surveys such as the Dark Energy Survey (DES) 
\footnote{http://www.darkenergysurvey.org},
or in the near future HETDEX \citep{HETDEX}, 
DESI \citep{DESI}, Euclid \citep{Euclid} and LSST \citep{LSST} this will also require
a large suit of mock galaxy catalogues. 

Ideally a set of 
high resolution N-body cosmological simulations, including hydrodynamics, would be run to generate
the mock galaxy catalogues, but in practise the computational time that this would require is 
exceedingly expensive. For this reason other methods to generate a large number of galaxy
mock catalogues quickly have been developed. 

\cite{Manera13}, inspired by \cite{Sco02}, developed the PThalos method to generate fast 
mock galaxy catalogues. The two main steps are: i) generate a matter field using 
Second Order Lagrangian Perturbation Theory (2LPT) ii) populate the field with halos and
galaxies, using a prescription that is calibrated against numerical simulations,
and that reproduces the observed clustering of galaxies (see Section \ref{sec:method} for details)
This method was used to create mock catalogues for the BOSS CMASS DR9 sample, which were
used in analyses by  
\cite{Anderson2012,Tojeiro:2012,Samushia:2012,Aardwolf2013,Reid2012,Sanchez2013}.

Other methods have also been developed to generate fast galaxy mocks. 
\cite{Monaco13,Monaco02} uses the collapsing times of dark matter particles in the Lagrangian field
smoothed at several scales to fragment the dark matter field into halos, giving clustering results 
similar to that of PTHalos. 
It has also been suggested that N-body simulations could be run with low time resolution. These runs
are 2-3 times slower than the methods based on a single field, but still at least 
within an order of magnitude or two faster than simulations. \cite{TassevCOLA}, 
suggested using 2LPT analytically at large scales, and \cite{QPM}, advocate essentially 
to run a Particle-Mesh simulation in this way. They obtain an increasing 
accuracy on the mass function and clustering with respect to the perturbation theory based methods.  

Finally, in order to increase the mass range of the halos from which galaxies are drawn in mocks, one can
use the conditional bias of halos as a function of the local halo density \citep{delaTorre13} 
use the halo mass function as a function of local mass density from a smaller size 
but higher resolution N-body simulations \citep{Angulo13}, or alternatively combine the 
2LPT approach for large scales with the Spherical Collapse model for small scales \citep{Kitaura13}.
We do not apply such methods here and
use a mapping between the mocks and corresponding N-body simulations instead
(see Section. 3 for details).    
 
In this paper we present 1000 mock galaxy catalogues for the LOWZ DR10 and DR11 BOSS galaxy
samples. These mocks were produced using the PTHalos 
method developed in \cite{Manera13} but with 
a redshift dependence in the Halo
Occupation Distribution (HOD) of galaxies in halos in order to account for the change of the
galaxy number density in redshift. The mocks are fitted to the LOWZ sample clustering presented in 
\cite{Tojeiro2013}
and the masks applied to the mocks mimic the survey geometry, observational completeness, and 
small-scale features such as patches of bad imaging and bright stars. 
These mocks have been used to provide covariance matrices and enable the study
the systematic and statistical uncertainty on the BAO scale measurements presented in   
\cite{Aardwolf2013,Sanchez2013,Chuang13}.
The mocks will be made publicly available.\footnote{www.marcmanera.net/mocks/}  

This paper is organized as follows. 
In Section 2 we introduce the LOWZ sample. We summarise the PThalos method and discuss the
geometry of the sample and the masks in Section 3. The HOD fitting is presented in Section 4. 
In Section 5 we explain the results and conclude in Section 6. Finally, tables with the covariance
matrices of the LOWZ sample correlation function are provided.

\section{BOSS LOWZ galaxy sample}

\begin{figure*}
\begin{center}
\includegraphics[width=3.3in]{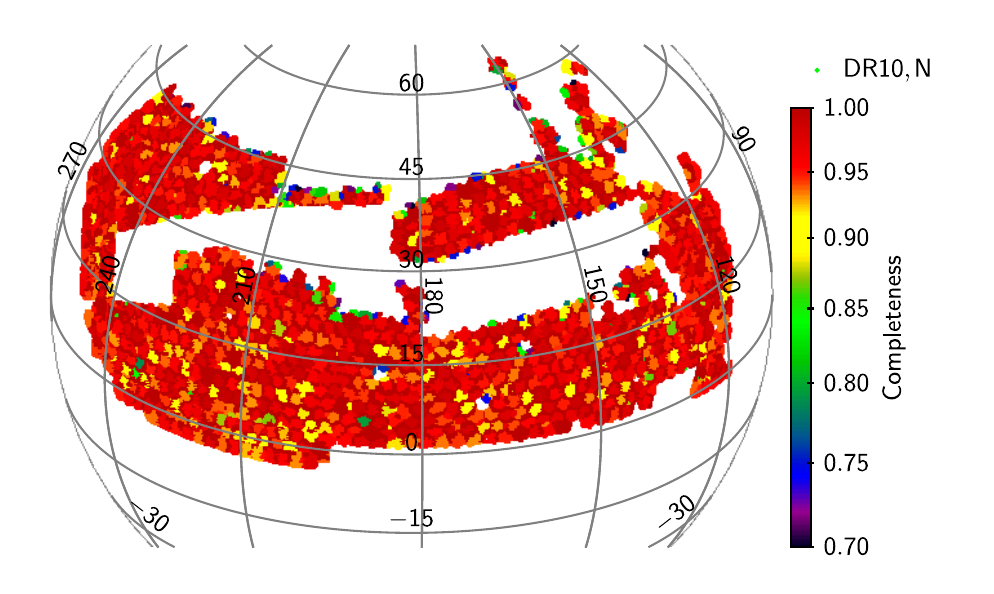}
\includegraphics[width=3.3in]{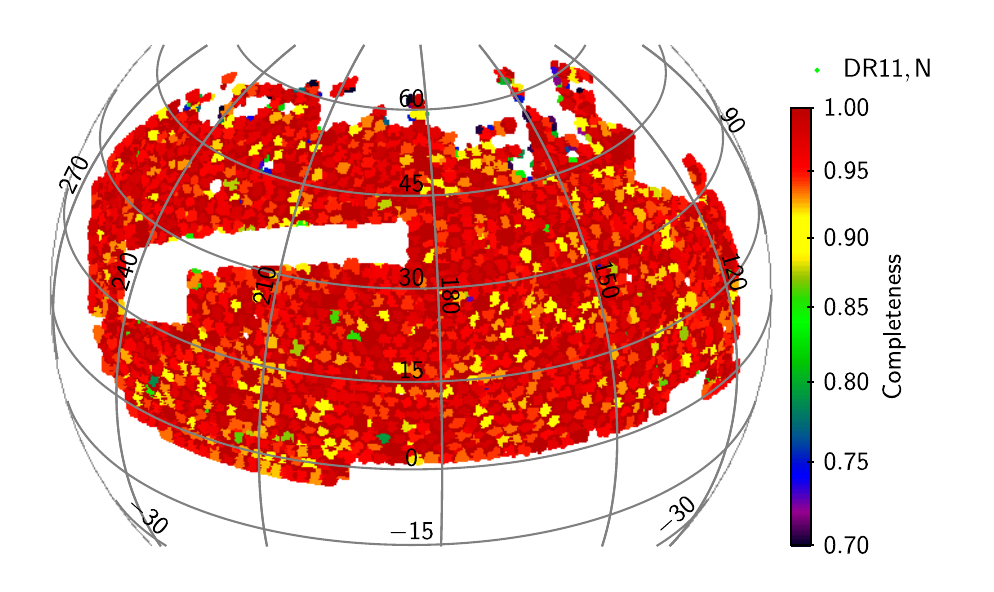}
\includegraphics[width=3.3in]{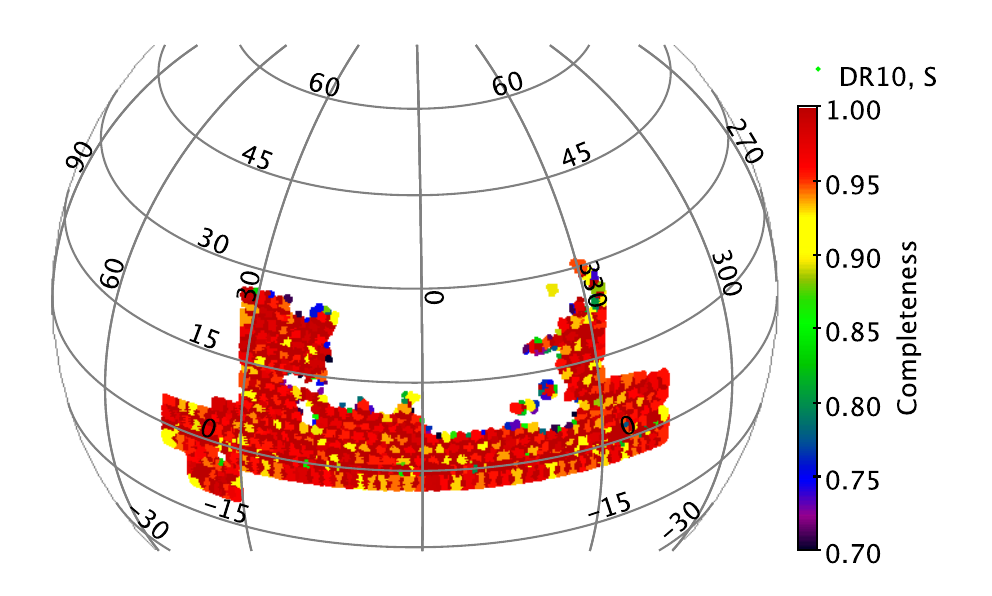}
\includegraphics[width=3.3in]{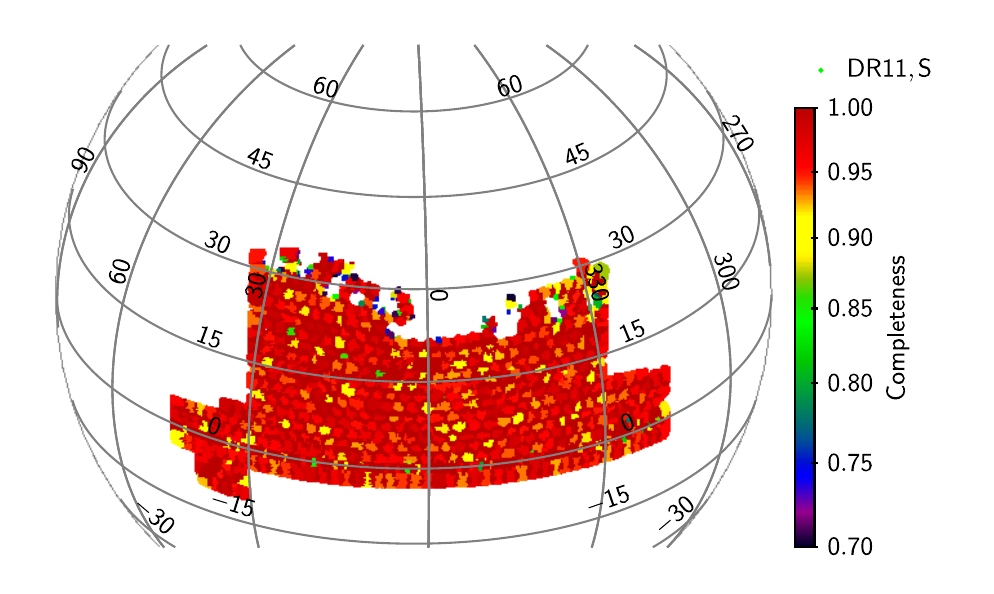}
\caption{DR10 (left) and DR11 (right) survey footprints in equatorial coordinates.
Northern Galactic Caps on the top and Southern Galactic caps on the bottom. The color code shows the completeness of each sector.} 
\label{footprints}
\end{center}\end{figure*}

BOSS uses SDSS CCD photometry (Gunn et al. 1998,2006) from five passbands (\textit{u, g, r, i, z}; e.g.,
Fukugita et al. 1996) to select targets for spectroscopic observation.
The LOWZ galaxy sample of BOSS targets galaxies with an algorithm  
that follows closely the one designed in SDSS-I/II  
for Luminous Red Galaxies, 
described in \cite{Eis01}, but extending to fainter 
magnitudes to increase the number density. This data set has already been used by 
\cite{Aardwolf2013,Parejko13,Tojeiro2013}. 

The LOWZ Data Release 10 (DR10), covers a total area of $5,635\,$deg$^2$  and is split in
two separate contigous regions. One is in the Northen Galactic Cap (NGC) and covers $4,205\,$deg$^2$. 
The other is in the Southern Galactic Cap (SGC) and covers $1,430\,$deg$^2$. The total LOWZ DR10 sample
has $218,905$ galaxies with $0.15 \lesssim z \lesssim 0.43$. The DR11 covers
respectively a total area of $7,998\,$deg$^2$  ($5,793\,$deg$^2$ in the NGC and $2,205\,$deg$^2$ in the SGC)
and has a total of $313,780$ galaxies. The NGC footprint is smaller than that of the CMASS
sample as the final target algorithm was not used for LOWZ 
during the first nine month of BOSS observations.
The footprint of the LOWZ sample is shown in Figure \ref{footprints}. 

\cite{Parejko13} studied the small scale clustering of the LOWZ sample and showed that 
the LOWZ galaxies occupy halos of typical mass of $\sim 5 \times 10^{13}$h$^{-1}$M$_{\odot}$, 
and galaxy bias $b\sim 2.0$. The large scale clustering of the LOWZ sample is 
presented in \cite{Tojeiro2013}, where the observational
systematics of the sample are studied in detail. The effective isotropic distance at $z=0.32$, 
$D_V = [ cz (1+z)^2 D_A^2/H ]^{1/3}$, where $H$ is the Hubble parameter and $D_A$ the angular
diameter distance, has been measured using the LOWZ BAO peak with an accuracy 
better than  2 per cent.
The cosmological implications of this measurement when combined with the BAO measurement
from the CMASS sample are presented in \cite{Aardwolf2013}. Both papers have used the PTHalos mocks galaxy catalogues
for the covariance matrices and analysis of errors.

\section{Method}

\label{sec:method}

We have created 1000 mock galaxy catalogues for the LOWZ DR10 and DR11 galaxy samples. 
We use a method similar to that of \cite{Manera13}, adapted to 
a lower redshift and with several improvements. The mocks are such that covariance matrices 
can be computed and the methods of analysis of the galaxy clustering can be tested for bias 
and relative accuracy. The steps that we took in generating these PTHalos mocks can be 
summarised as follows:

\subsection{Dark matter}

We have run 500 dark matter particle fields based on  2LPT, 
using the pubicly available code 2LPTic\footnote{http://www.marcmanera.net/2LPT/}.
The matter fields were generated at redshift $z=0.32$, which is the effective pair-weighted
redshift of the LOWZ sample. The matter realizations have been generated in a cubical box of 
size $L=2400$ h$^{-1}$ Mpc with $N=1280^3$ particles, for a $\Lambda$CDM cosmology with parameters 
$\Omega_m = 0.274$, $\Omega_\Lambda=0.726$, $\Omega_b = 0.04$, $h=0.7$,
$\sigma_8=0.8$ and $n_s=0.95$, giving a particle mass $M_{p}=50.1\times 10^{10} M_{\odot}/h$
The input power spectrum has been smoothed with a 
cut-off as in \cite{Manera13} as it helps the clustering of small halos. 
These cosmological parameters are the same as the standard fiducial choices used in
BOSS analyses.

\subsection{Halos}

Halos have been identified in the dark matter field using a Friends-of-Friends halo-finder algorithm 
\citep[FoF,][]{DEFW}
which percolates in a halo all the particles that can be linked by within a given linking-length $l$. 
The value of the linking-length used in N-body simulations varies in the literature, the most
common value being $l=0.2$ times the mean inter-particle distance. For a 2LPT dark matter field the value of the 
linking-length has to be appropriately changed in accord with the 2LPT dynamics. In the spherical collapse 
approximation both values can be related as follows \citep{Manera13}:

\begin{equation}
l_{2LPT} = b_{N-body} \left( \frac{\Delta_{vir}^{N-body}}{\Delta_{vir}^{2LPT}} \right)^{1/3} \; , 
\label{linkL}
\end{equation}
where $\Delta_{vir}^{N-body}$ and $\Delta_{vir}^{2LPT}$ are the N-body and 2LPT virial overdensities.
For the N-body halos we take the value of Bryan and Norman (1998) fit,
\begin{equation}
\Delta_{vir}^{sim}=[18\pi^2+82(\Omega_m(z)-1)-39(\Omega_m(z)-1)^2] /\Omega_m(z),
\label{BryanNorm}
\end{equation}
where $\Omega_m(z)=\Omega_m (1+z)^3 H(0)/H(z)$, and $H(z)$ is the Hubble expansion value at redshift $z$.
For the 2LPT halos we take $\Delta_{vir}^{2LPT}=35.4$. This value comes from the relation between
the linear and non-linear density in 2LPT,
\begin{equation}
\Delta_{vir}^{2LPT}=\delta_{NL}^{2LPT}+1= (1- \delta_0 D_1 /3 + \delta_0^2 D_1^2/9)^{-3} \,,
\end{equation}
where 
$\delta_0 D_1=1.868$ is the value of the linear density fluctuation at collapse, and $\delta_{NL}$, 
its non-linear value. 

Applying the above equations, we use the linking length value $l=0.388$ for halos at redshift $z=0.32$. 
As the value is approximate, and 2LPT lacks small scale power, the halo mass function recovered would
not completely match that of an N-body simulation. Therefore, following \cite{Manera13}, we re-assigned 
the masses of the halos, while keeping their positions and rank-order in mass, 
such that we recover the \cite{Tin08} mass function for this cosmology. This method has been 
shown to match the clustering of halos in N-body simulations within 10 per cent accuracy. 

\subsection{Galaxies}

We assign galaxies into halos by means of an Halo Occupation Distribution (HOD; Peacock \& Smith 2000, Scoccimarro et al. 2001, Berlind \& Weinberg 2002) functional form with five parameters, as used by Zheng et al. (2007). The mean number of galaxies in a halo of mass $M$ is the sum of the mean number of central galaxies plus the mean number of 
satellite galaxies, $\langle N(M) \rangle = \langle N_{cen}(M)\rangle + \langle N_{sat}(M)\rangle$, where 

\begin{eqnarray}
\langle N_{cen} \rangle  &=& \frac{1}{2} \left[ 1 + {\rm{erf}}\left( \frac{{\rm log} M - {\rm log} M_{cut} }{\sigma_{{\rm log} M}}\right)\right]  
\nonumber \\
\langle N_{sat}\rangle &=& \langle N_{cen}\rangle \left( \frac{M-M_0}{M_1}\right)^\alpha \; , 
\label{hodZeng}
\end{eqnarray}
and $\langle N_{sat}\rangle = 0$ if the halo mass has $M < M_0$. 
In this parametrization $M_{cut}$ and $M_1$ are the typical halo masses
for having respectively order of one central and one satellite galaxy\footnote{ In this paper \textrm{log} always stands for base-10 logarithm.}. 
The HOD parameters are calibrated to fit the observational data (see Section \ref{sec:fitting}). 
Galaxies in halos are given the velocity of the halo, plus a dispersion velocity
from a Gaussian distribution with an amplitude given by the mass of the halo
and the Virial theorem. Galaxies that are below our lower halo mass 
limit of $5 \times 10^{12} M_{\odot}/h$ (~7 per cent of the total) 
are assigned randomly to dark matter particles that 
do {\it not} belong to halos. This is different from the CMASS mocks in \cite{Manera13}
where we randomly assigned these galaxies to any dark matter particle.  
More importantly, we have allowed the HOD to depend on the number density of galaxies, and
fitted the HOD {\it simultaniously} with the number density as a function of redshift, 
therefore, we have {\it not} subsampled the galaxy field a posteriori to match the LOWZ distribution.   
The details of the fitting procedure are explained in Section \ref{sec:fitting}.

\subsection{Mask: geometry}

\begin{figure}
\center
\includegraphics[width=70mm]{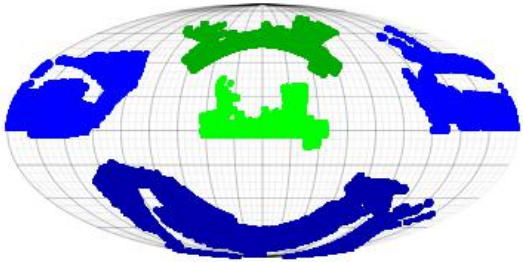}
\caption{Footprints of the LOWZ DR10 NGC and SGC mock galaxy catalogues. Two of each can fit without overlap in the celestial sphere. The same is true for the DR11 footprints.}
\label{fig:area12}
\end{figure}

\begin{table}
\begin{tabular}{lrr}
\hline
\hline                                            
LOWZ DR10 & NGC  &  SGC \\                       
\hline                                          
Total area / deg$^2$ &     4222 & 1429    \\       
Veto area  / deg$^2$ &      251 &   58    \\    
Used area  / deg$^2$ &     3971 & 1371    \\   
Effective area / deg$^2$ & 3840 & 1331    \\   
\hline
\hline 
LOWZ DR11 & NGC  &  SGC \\
\hline
Total area / deg$^2$ &  5787 & 2204 \\                                 
Veto area  / deg$^2$ &   335 &   89 \\            
Used area  / deg$^2$ &  5452 & 2115  \\        
Effective area  / deg$^2$ & 5287 & 2060 \\     
\hline
\hline
\end{tabular}
\caption{Areas of the LOWZ sample mock galaxy catalogues.}
\label{tableareas}
\end{table}

BOSS observes regions of the sky in the two galactic hemispheres. 
Figure \ref{footprints} shows the NGC and the SGC 
observed footprints for the LOWZ Data Release 10 and 11 (DR10,DR11),
the latter more than double the areas observed by BOSS in DR9. 

As with the data, the footprints of the mock galaxy catalogues exclude vetoed regions,
which amounts to about 5 per cent of the total area covered. 
These regions are generally small and have been removed for a variety of reasons including regions with bad photometry, failing of the PSF modelling, timing out errors in the pipeline reduction, 
or regions around bright stars, or around objects that have been highly prioritized, since a galaxy
cannot be observed within the fibre collision radius of these points. For more detailed 
information of the veto mask see \cite{Anderson2012} and SDSS DR10 documentation. 

In table \ref{tableareas} we show the areas of the NGC and SGC of the LOWZ DR10 and DR11 mock
galaxy catalogues. There are small differeences (less than 0.5 per cent) between the areas of
the mocks and of the data, which result because of "last minute" changes to the data mask used. 
The effect of these differences is insignificant. The effective area is the area used weighted
by the target completeness.  

Regarding the geometry of the LOWZ sample, it is worth noticing that it is possible
to fit two samples of the NGC and SGC footprints in the celestial sphere without overlap.
We have taken advantage of that when creating our mock galaxy catalogues. In this way
we only needed 500 matters field to generate 1000 mocks. 
In order to get two footprints within the same matter run, we convert 
the right ascension, $\rm{ra}$, 
and declination, $\rm{dec}$, to cartesian coordinates and then rotate about the $y$ axes
using the standard transformation between $\rm{ra}$ and $\rm{dec}$ and the cartesian coordinates
\beqa
\hspace{2cm} x & = & \rm{cos}(\rm{ra}) \, \rm{cos}(\rm{dec}) \nonumber \\
\hspace{2cm} y & = & \rm{sin}(\rm{ra}) \, \rm{cos}(\rm{dec}) \nonumber \\
\hspace{2cm} z & = & \rm{sin}(\rm{dec})\, . 
\eeqa
The vector $r={x,y,z}$ can be easily rotated by an angle $\alpha$ by $r'=\text{Ry}[\alpha]\,r$,
with the matrix of rotation about the $y$ axis  
\begin{equation}
\text{Ry}[\alpha ]=\left(
\begin{array}{ccc}
 \text{cos}[\alpha ] & 0 & \text{sin}[\alpha ] \\
 0 & 1 & 0 \\
 -\text{sin}[\alpha ] & 0 & \text{cos}[\alpha ]
\end{array}
\right)\; .
\end{equation}

Figure \ref{fig:area12} shows two NGG and two SGC footprints. The second footprints of the 
NGC and SGC are obtained by rotating the previous ones respectively with 
$\alpha=-120$deg and $\alpha = -55$deg.

\subsection{Mask: completeness}

The mocks have been created taking into account the completeness of the sample 
observed at every sector in the sky, as measured from the data. We do not 
re-position plates for each mock as if we were performing actual observations. 
The mock galaxies have been subsampled
to mimic variations in the target completeness, redshift failures, and close pair completeness. 
Close pair completeness refers to the case where a spectroscopic redshift of a galaxy
is not available due to the fact that is is within 62$''$ of another galaxy, meaning
that two fibers cannot be placed on both galaxies simultaneously. 
The effective areas of the mocks, that result from weighting by a measure of target
completeness, $C_{BOSS}$, as defined in \cite{Anderson2012} are shown in table \ref{tableareas}.
For detailed numbers of galaxies, missed targets and areas of the LOWZ galaxy 
sample see \cite{Tojeiro2013}.

\section{Modelling the galaxy distribution}
\label{sec:fitting}

\subsection{HOD(z)}

\begin{figure}
\center
\hspace{-20pt}\includegraphics[width=77mm]{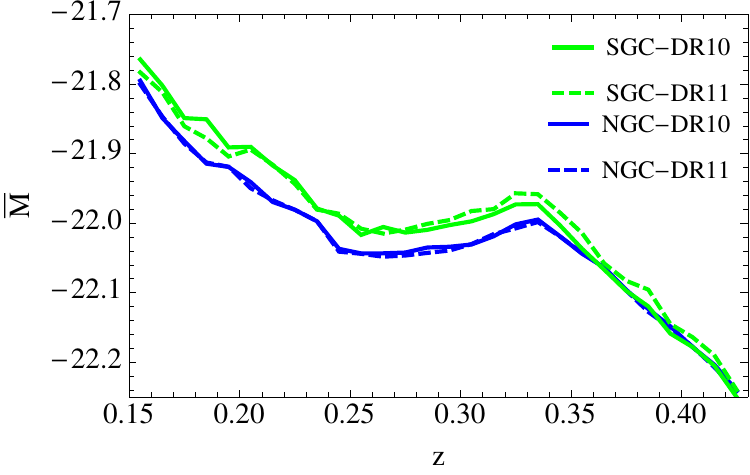}
\includegraphics[width=70mm]{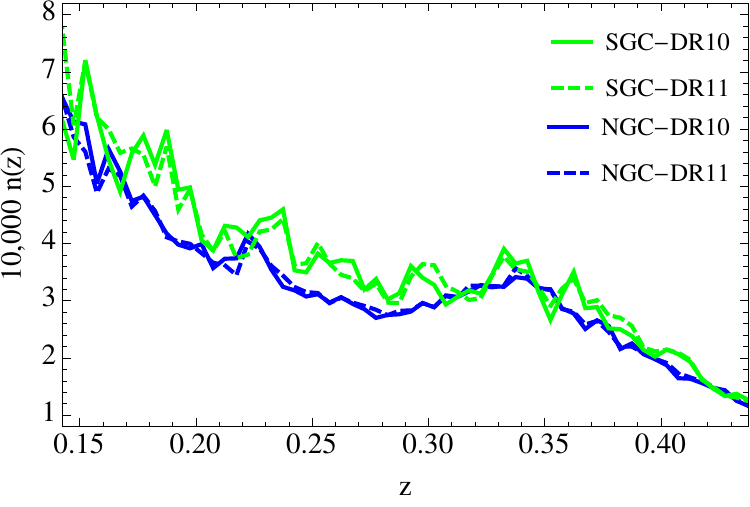}
\caption{Top: Average of the absolute magnitude of the LOWZ DR10 (solid line) and DR11 (dashed line) galaxy samples. 
Bottom: Number density of galaxies of the LOWZ DR10 (solid line) and DR11 (dashed line) galaxy samples.
In both panels the blue lines show the NGC sample and the green lines the SGC.} 
\label{ndz_plot}
\end{figure}

Given a Halo Occupation Distribution set of parameters the number density of galaxies is fixed;
it can be obtained as an integration of the halo mass function weighted by the HOD. Previous
mock galaxy catalogues based on populating halos from a cosmological time-slice 
by means of an HOD have, by construction, a constant number density of galaxies. Consequently,
to mimic the number density as a function of redshift, the number 
of galaxies must be subsampled a posteriori \citep{Xu13,Manera13,McBride13}. 
Randomly subsampling a distribution of galaxies would not change any of its fundamental 
properties apart from the number density itself. 

In the top panel of Figure 
\ref{ndz_plot} we show the
average absolute magnitude of the (k-corrected r-band) galaxies in the LOWZ sample for 
DR10 and DR11. We see that the average magnitude of the sample varies with redshift for both 
the NGC or SGC. Moreover the shape as a function of redshift is similar 
to that of the number density, which we show in the bottom panel of 
Figure \ref{ndz_plot}. This suggests
that the HOD parameters are \textit{not} likely to be well approximated as constant with redshift.

In this paper we want to improve on the mocks by allowing the HOD parameters to vary as a function 
of redshift. Fitting a different HOD pareametes in redshift slices would not constrain 
the HOD parameters sufficiently, so we have chosen instead to include the redshift dependence through
a fixed dependency of the HOD parameters on the number density of galaxies. 
This dependency is based on previous studies, as we now describe.

\begin{figure}
\center
\includegraphics[width=70mm]{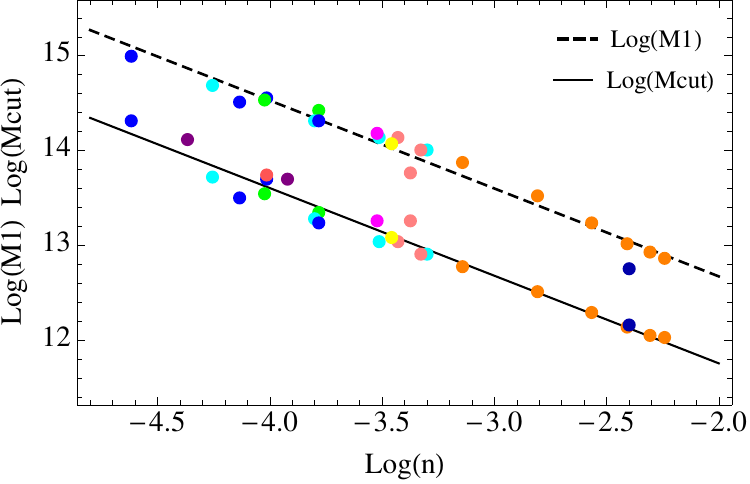}
\caption{HOD parameters $M_{cut}$ and $M_{1}$ as a function of the number density of galaxies.
The points are from the list of table A1 in \protect\cite{Parejko13}. Blue: SDSS LRG (2PCF), \protect\cite{Zheng09};
\protect\cite{Mandelbaum06}; Cyan: SDSS LRG (Photo-z, BCL), \protect\cite{Blake08};
Pink: SDSS LRG (Photo-z, PW), \protect\citet{Padmanabhan09}; Dark Blue: Combo-17
\protect\cite{Phleps06}; 
Purple: SDSS LRG (Lensing), \citet{Mandelbaum06}; Red: SDSS LRG (3PCF),\citet{Kulkarni07};
Orange: NDWFS, \protect\cite{Brown08};
Green: 2SLAQ \protect\citet{Wake08}; \protect\cite{Kulkarni07}; Yellow: BOSS CMASS, \protect\cite{White11};
Magenta: LOWZ \protect\cite{Parejko13} }
\label{fig:hodnz}
\end{figure}

\cite{Parejko13} report a compilation of the Mcut and M1 HOD parameters from various papers in the literature based on a variety of different galaxy samples. While the functional form of the HOD used in these papers varies, they are sufficiently similar to allow for a comparison between the derived parameters and thus study the evolution of the HOD. We have used the data from Table A1 in \cite{Parejko13}to fit a log-linear dependence of Mcut and M1 as a function of the number density of galaxies used in each paper.
In Figure \ref{fig:hodnz}, we show the data and our best fits,  

\begin{eqnarray}
\textrm{log} M_{cut} & = & \textrm{log} M_{cut}^0 + S_{cut} \bar{n} \; , \nonumber \\ 
\label{eq:SS}
\textrm{log} M_{1} & = & \textrm{log} M_{1}^0 +  S_1 \bar{n}  \; ,\\ \nonumber 
\end{eqnarray}
where $\textrm{log} M_{cut}^0 = 9.90 \pm 0.12$, $ S_{cut} = -0.925 \pm 0.035 $,  
$\textrm{log} M_{1}^0 = 10.81 \pm 0.12$ and  $S_1 = 0.928 \pm 0.037$.

We have considered data from publications that include the parameters
$\alpha$, $\kappa=M_0/M1$, or $\sigma_{logM}$, and found no significant 
dependency of these parameters on the number density of galaxies.  
Consequently, when fitting the HOD parameters for our mocks, we keep
these parameters constant as a function of redshift, while for 
$M_{cut}$ and $M_1$ we have fixed the tilts $S_{cut}$ and $S_1$ to
the best fit values given the previous data and fitted only the amplitudes $M^0_{cut}$ and $M^0_1$
to the BOSS data. For a redshift and luminosity depenence of the HOD see also \cite{Zheng07,Coupon12,Tinker13,Hong14}.

\subsection{Fit to n(z)}

The actual number density of observed LOWZ galaxies varies as a function of redshift 
for two main reasons. The principal effect is due to the color and magnitude cuts
of the target selection that induce a smooth redshift dependence. In addition, there
are 'high-frequency' variantions in redshift that come from observing a particular volume  
of the universe, i.e, cosmic variance. The shot-noise contribution from being a sample with 
a finite number of galaxies is subdominant respect to the cosmic variance.

We creating our suit of mock catalogues, we aim to an average redshift distribution that 
matches the smooth component of the observed redshift profile without the noisy component
that is specific to the observed sample. The noisy contribution is accounted for as 
each mock is a different realization of our universe, within our fiducial cosmology.  

We have therefore smoothed the observational galaxy redshift distribtuion to obtain the target n(z)
to which we fit the HOD of the mock galaxy catalogues. The smoothed n(z) is a cubic spline
curve with seven nodes. The number of nodes and their n(z) values have been determined with
a minimization process. First, we have estimated a covariance matrix of n(z), in bins of
0.05, from a preliminary version of the mocks that already included a redshift dependent 
number density. Then, using this covariance matrix, we have fitted a set of cubic splines 
to the observed n(z), each spline with a different number of nodes. For each
of these splines we have set the n(z) values by minimising the $\chi^2$ 
against the observed n(z). As expected the goodness of fit increases with the number
of nodes but at the expense of mimicking all the little wiggles that are inducced
by cosmic variance. Consequently, we have used the lower number of nodes that 
fit the data with $\chi^2 \sim 1$ per degree of freedom. 

We have found that, for the NGC, seven nodes between $0.1175 < z < 0.4425$ fit 
well the redshift distribution, so we have used this number for our n(z) spline. 
The redshift range is broader than the one we use for our LOWZ sample $0.15 < z < 0.43$
to allow for redshift space distortions that may cross the redshift boundary. 
We have fitted the SGC with the same number of nodes, as we expect the smooth 
component of n(z) to be similar in the two hemispheres and the NGC measured n(z)
has a higher signal-to-noise (see \citep{Tojeiro2013} for a discussion of NGC and SGC
differences).

\begin{figure}
\center
\includegraphics[width=85mm]{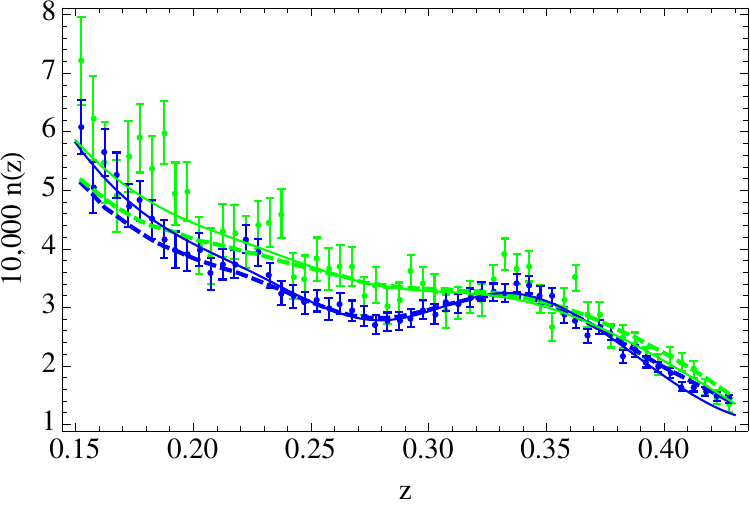}
\caption{Number density of galaxies of the LOWZ DR10 galaxy sample for the NGC (blue dots, lower values) and the 
SGC (green dots, higher values). Error bars show the rms of the 1000 mock galaxy catalogues. The solid lines are 
fits to the data. The mock galaxy catalogues n(z) are shown in dashed lines;
all the redshift dependence of the mocks n(z) comes through the variation of the HOD.}
\label{fig:nbar1}
\end{figure}


We have set the HOD parameters by minimizing the $\chi^2$ value of the power 
spectrum and the number density. The joint $\chi^2$ is thus the addition of 
the two respective contributions:

\begin{equation}
\begin{split}
\chi^2=\sum_{i,j} (P^m(k_i) - P^d(k_i)) C_{i,j}(P)^{-1} (P(k_j)^{m} - P(k_j)^{d}) \\ 
+ \sum_{s t} (n^m(z_s) - n^d(z_s)) C_{s,t}(n)^{-1} (n(z_t)^{m} - n(z_t)^{d}) ,  
\end{split}
\label{eqchi}
\end{equation}
where $P(k_l)$ is the value of the power spectrum at wave number bin $l$, and
$C_{i,j}(P)$ is the covariance matrix of the power spectrum. In the
same manner $n(z_l)$ is the value of the number density at redshift bin $l$, and 
$C_{s,t}(n)$ the covariance of the number density of galaxies. The
labels $d$ and $m$ stand for data and mocks. 
We have fitted $P(k)$
in the range $0.02 < k < 0.15$ and $n(z)$ in the range $0.15 < z < 0.43$. 
For each HOD set of parameters
that we have run we took the mock power spectrum and number density to be the mean of  
ten realizations for the NGC and twenty for the SGC.  In this way we reduce 
the effect of fiting the data with only one mock catalogue.
We have used twenty mocks for the SGC and then for the NGC in order to have a similar
number of galaxies in both cases. 

To minimize the $\chi^2$ we have used the simplex algorithm of Nedler and Mead (1969).
This method constructs a multi-dimensional simplex with vertices given by the initial guess 
of the HOD parameters and a certain step-size. By a series changes of the position
of the vertex with worst $\chi^2$ the simplex moves in the parameter space until
it brackets the minimum within a given volume. 

For the covariance matrices we have used an estimation of a preliminary version
of the mocks that had been created in the same manner starting with the HOD
parameters of \cite{Parejko13}. With this covariance matrix we then minimized
the HOD, separately for the NGC and SGC, obtaining the following best fits: \\ 
\begin{center}
\begin{tabular}{lrr}
\hline
\hline
param & NGC & SCG \\
\hline
$\textrm{log} M_{cut}$ & 13.20 & 13.14 \\ 
$\textrm{log} M_1$ & 14.32 & 14.58 \\     
$\textrm{log} M_0$ & 13.24 & 13.43 \\     
$\sigma_{logM}$ & 0.62 & 0.55 \\          
$\alpha$ & 0.93 & 0.93 \\                
$\chi^2$ & 49  & 30 \\
\hline\hline
\end{tabular}
\begin{tabular}{c}
{\small \textbf{Table 2.} HOD values for the LOWZ mock catalogues.} 
\end{tabular}
\end{center}
where $\textrm{log} M_{cut}$, $\textrm{log} M_1 $ and $\textrm{log} M_0$ are the values of these
parameters when in Eq (\ref{eq:SS}) we set $n(z)=2.98\cdot 10^{-4}$ and  $S_{cut}$ and $S_1$ 
are respectively $-0.925$ and $-0.928$.

As the HOD that we are using has five free parameters there is some room
for the best fit to vary depending on the initial guess at which the fiting starts 
as well as the particular set mock realizations used to fit the data. We also expect
the observational HOD to be different due to the fact that 7 per cent of our galaxies  
are not in resolved halos in our simulation. 
The values we have found for the HOD parameters are within one sigma of the mean of the full sample  
in \cite{Parejko13}. The recovered $\chi^2$ values for our best-fit HOD models show they are a
good fit to the data. Indeed, since we have 88 bins in total (32 from $P(k)$ and 56 from $n(z)$), 
the $\chi^2$ square values are less than the number of degrees of freedom, and thus a 
good fit for the purposes of creating mocks for covariance matrices and clustering data analysis.  

In Figure \ref{fig:nbar1} we show the number density of galaxies in 
the DR10 LOWZ sample for the NGC and the SGC, with errors displaying 
the rms of the 1000 mock catalogues. The solid lines show the mean 
of the targeted n(z) that comes from the seven-node spline fit to the data, 
and the dashed lines shows the mean n(z) of the mock galaxy catalogues.
The number density of the mocks have not been subsampled and its 
redshift dependence comes only through Eq (\ref{eq:SS}) after
fitting for the HOD parameters. We recover  
the redshift distribution for $z > 0.2$ quite well. At lower redshift the
differences come from the fact that the log-linear (or constant) 
dependence of the HOD mass parameters as a function of n(z) is 
an aproximation to the true HOD as function of redshift.   

Following the methods outlined in Sections \ref{sec:method} and \ref{sec:fitting}  
we have created one thousand mock galaxy catalogues
for the LOWZ DR10 and DR11 galaxy sample. Since both releases have the same targeting their clustering
and redshift distributions are very similar. We have consequently used the same halo fields and HOD 
parameters for both releases, those fitted with DR10 data.

\section{Results}

\begin{figure*}
\centering
  \includegraphics[width=.49\linewidth]{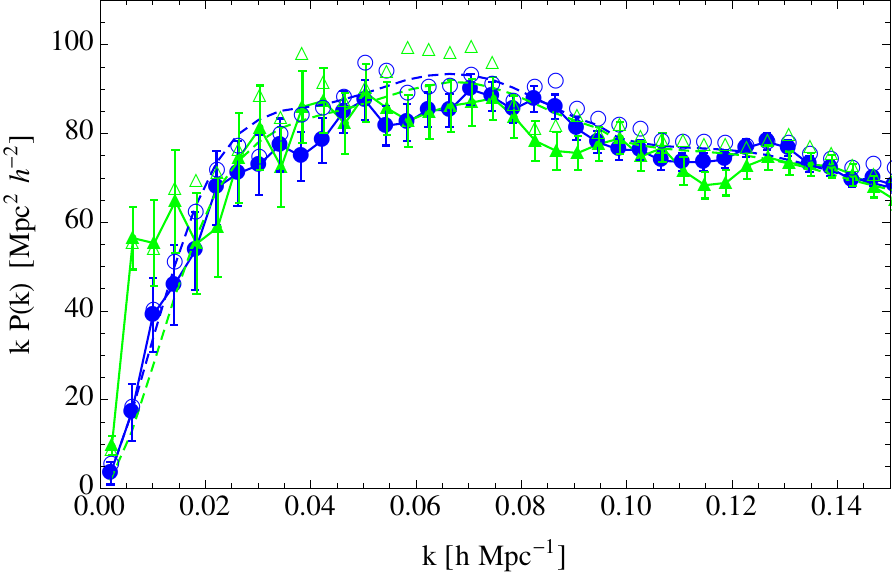}
  \includegraphics[width=.49\linewidth]{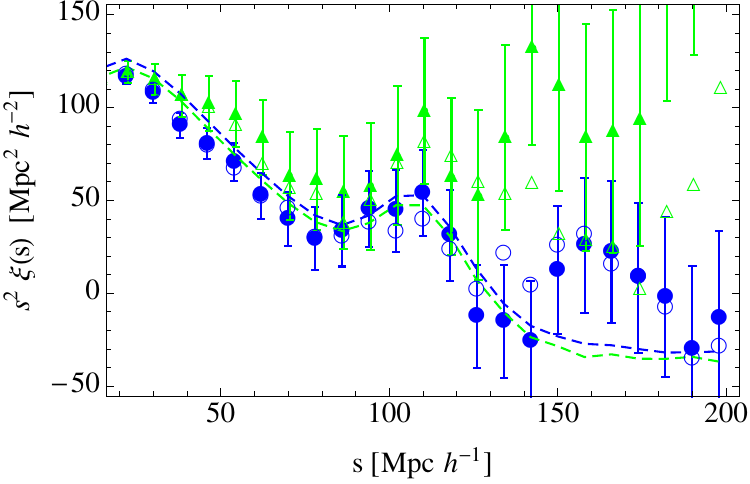}
  \caption{
Power spectrum and correlation function of the LOWZ galaxy sample. The DR10 NGC and
SGC measurements are shown as blue solid circles and green solid triangles respectively, 
with sample errors from the dispersion of the mock galaxy catalogues. DR11 clustering 
is displayed in open symbols. The mean values of the mock catalogues are shown for the NGC and SGC as blue and red 
lines. 
}
  \label{fig:mocks}
\end{figure*}

In the left panel of Figure \ref{fig:mocks} we show the 
power spectrum of the DR10 LOWZ galaxy sample with errors from the mocks, both for the
NGC (blue solid circles) and the SGC (green solid circles). The DR11 values are set as open symbols. 
We have used the \cite{fkp94} (FKP) estimator. The mean of the mock catalogues is shown 
by the solid lines. There is a good fit between the data and the mock catalogues for $k > 0.02$,
which is the region in which we have fitted the HOD. At lower $k$ values the power 
of the mock catalogues decreases, as expected for any $\Lambda$CDM cosmology with 
typical values from WMAP or PLANCK measurements. The power of the
DR10 galaxy sample for the NGC decreases as well, but the SGC increases
having extra power compared to the NGC. 
\cite{Tojeiro2013} have looked at the differences between NGC and SGC data 
in terms of systematics and found that none of the systematic contributions
analysed in the CMASS sample has an significant impact in the LOWZ data.
\cite{Tojeiro2013} found that the differences between the two galactic caps are reduced in DR11
and are compatible with one another, given the expected variance computed from the mocks. 

In the right panel of Figure \ref{fig:mocks} we show the two-point correlation function 	 
of the DR10 LOWZ galaxy sample both for the NGC (blue solid circles) and SGC (green solid). The
DR11 values are set as open circles. We
have used the \cite{lansza93} estimator, including FKP weights as it reduces the variance
also in the correlation function. The error
bars are the sample errors computed from the mock catalogues and the values 
of the mean of the mocks are shown as solid lines. 
The excess power of the SGC at low $k$ translates into a higher values of the
correlation function at a wide range of scales. 

We present the values of the DR10 LOWZ power spectrum and correlation function, 
and their covariance matrices in the appendix of the paper. The mocks for LOWZ 
DR10 will be made publicly available online.
\footnote{www.marcmanera.net/mocks/}
These mocks have been used by the BOSS collaboration in analysis of the large 
scale of the LOWZ sample and the BAO peak position \citep{Aardwolf2013,Tojeiro2013,Sanchez2013}. 
The elements of covariance matrices estimated from a finite set of mocks 
have uncertainties that depend on the number of mocks. These uncertainties 
translate into errors in the inverse covariance matrices and likelihood estimators. 
Since our suit consists of 1000 mocks these errors
are expected to be small in most cases, and are of order few percent for 1000 mocks
using $~30$ bins. For a detailed accounting for this errors see \cite{Percival13}.

\section{Conclusion}

We have created 1000 mock galaxy catalogues for the BOSS LOWZ DR10 and 
DR11 galaxy sample. These mock catalogues have been produced following
the PTHalos method developed in \cite{Manera13}, but with significant differences.  
We have created 500 particle dark matter fields using a 2LPT code and obtained halos 
in those fields by FoF method with the appropriate linking length. The mass of the halos have been ranked and 
masses re-assigned to match the \cite{Tin08}. These PTHalos have been populated with
galaxies. For each matter field we can fit two full LOWZ footprints without overlap, 
resulting in a 1000 mocks for both the Northern Galactic Cap and the Southern Galactic Cap. 
Redshift space distortions are included through peculiar velocities. 

In contrast to previous mocks these  have
been created allowing for a variable HOD as a function of redshift, authomatically matching the number density
of galaxies. The mocks were created by fitting simultanously the measured clustering and number density,
with no need for applying a posterior subsampling of galaxies. We have implemented the DR10 and DR11
LOWZ masks to the mock catalogues, including small vetoed areas due to bright stars or other effects
like bad photometry and target completeness. We have also included close pair corrections, and redshift failures. 
For the fitting procedure and HOD dependence on number density see Section \ref{sec:fitting}. 
The one thousand LOWZ mocks galaxy catalogues have been used in the analysis of the Baryon Accoustic Peak position 
\cite{Aardwolf2013} and shape of the correlation function \cite{Sanchez2013}. 
In the appendix we present the LOWZ DR10 correlation function covariance matrix. 

Mock galaxy catalogues for the BOSS CMASS galaxy sample ($0.43 < z < 0.7$) have also been upgraded
from DR9 \cite{Manera13} to DR10 and DR11, keeping the same HOD but repopulating the halos and
applying the DR10 and DR11 footprints, completeness masks, and $n(z)$ fit as in Section \ref{sec:method}.
These mocks have been used in studying the clustering of red and blue galaxies,
\citep{Ross13} the accuracy of fitting methods \citep{Vargas13} and the analysis of
the the large scale clustering and its cosmological implications, including 
the BAO position, anisotropic clustering \citep{Aardwolf2013,Samushia:2013,Sanchez2013,Chuang13}.

The DR10 LOWZ and CMASS mocks will be publicly available online.$^5$ 

\section*{acknowledgments}

MM and WJP acknowledge support from European Research Council, through grant "MDEPUGS".  
MM is very thankful for Ramin A. Skibba and Bob Nichol useful comments and suggestions. 
Funding for SDSS-III has been provided by the Alfred P. Sloan Foundation, the Participating Institutions, the National Science Foundation, and the U.S. Department of Energy Office of Science. The SDSS-III web site is http://www.sdss3.org/.

SDSS-III is managed by the Astrophysical Research Consortium for the
Participating Institutions of the SDSS-III Collaboration including the
University of Arizona,
the Brazilian Participation Group,
Brookhaven National Laboratory,
University of Cambridge,
Carnegie Mellon University,
University of Florida,
the French Participation Group,
the German Participation Group,
Harvard University,
the Instituto de Astrofisica de Canarias,
the Michigan State/Notre Dame/JINA Participation Group,
Johns Hopkins University,
Lawrence Berkeley National Laboratory,
Max Planck Institute for Astrophysics,
Max Planck Institute for Extraterrestrial Physics,
New Mexico State University,
New York University,
Ohio State University,
Pennsylvania State University,
University of Portsmouth,
Princeton University,
the Spanish Participation Group,
University of Tokyo,
University of Utah,
Vanderbilt University,
University of Virginia,
University of Washington,
and Yale University.

Part of the numerical computations and analyses of this paper made use of the Sciama High Performance Compute (HPC) cluster which is supported by the ICG, SEPNet, and the University of Portsmouth; and of the COSMOS/Universe supercomputer, a UK-CCC facility supported by HEFCE and STFC in cooperation with CGI/Intel.

\appendix

\begin{table*}
\begin{tabular}{ccccccccc}
\hline
\hline
 \text{C(r1,r2)} & \text{   30} & \text{   38} & \text{   46} & \text{   54} & \text{   62} & \text{   70} & \text{   78} & \text{ 
    86} \\
\hline
 \text{   30} & \text{ 32.15} & \text{} & \text{} & \text{} & \text{} & \text{} & \text{} & \text{} \\
 \text{   38} & \text{ 23.15} & \text{ 21.58} & \text{} & \text{} & \text{} & \text{} & \text{} & \text{} \\
 \text{   46} & \text{ 15.69} & \text{ 15.86} & \text{ 15.52} & \text{} & \text{} & \text{} & \text{} & \text{} \\
 \text{   54} & \text{ 11.29} & \text{ 11.69} & \text{ 12.27} & \text{ 12.37} & \text{} & \text{} & \text{} & \text{} \\
 \text{   62} & \text{ 8.384} & \text{ 8.821} & \text{ 9.545} & \text{ 10.12} & \text{ 10.42} & \text{} & \text{} & \text{} \\
 \text{   70} & \text{ 6.186} & \text{ 6.786} & \text{ 7.479} & \text{ 8.015} & \text{ 8.698} & \text{ 8.879} & \text{} & \text{}
   \\
 \text{   78} & \text{ 5.013} & \text{ 5.452} & \text{ 6.112} & \text{ 6.504} & \text{ 7.085} & \text{ 7.511} & \text{  7.74} &
   \text{} \\
 \text{   86} & \text{ 4.142} & \text{ 4.437} & \text{ 4.936} & \text{ 5.268} & \text{ 5.718} & \text{ 6.068} & \text{ 6.576} &
   \text{ 6.736} \\
 \text{   94} & \text{ 3.112} & \text{ 3.244} & \text{ 3.691} & \text{  3.97} & \text{ 4.303} & \text{ 4.587} & \text{ 5.091} &
   \text{ 5.475} \\
 \text{  102} & \text{ 2.479} & \text{ 2.526} & \text{  2.88} & \text{ 2.978} & \text{ 3.251} & \text{ 3.542} & \text{ 3.976} &
   \text{ 4.306} \\
 \text{  110} & \text{ 2.118} & \text{ 2.019} & \text{ 2.175} & \text{ 2.223} & \text{ 2.419} & \text{ 2.674} & \text{ 2.967} &
   \text{ 3.209} \\
 \text{  118} & \text{ 1.544} & \text{ 1.516} & \text{  1.66} & \text{ 1.694} & \text{ 1.808} & \text{ 1.962} & \text{ 2.158} &
   \text{ 2.352} \\
 \text{  126} & \text{ 1.068} & \text{ 1.204} & \text{ 1.341} & \text{ 1.376} & \text{ 1.406} & \text{ 1.572} & \text{ 1.737} &
   \text{ 1.878} \\
 \text{  134} & \text{ 0.7244} & \text{ 0.8605} & \text{ 1.002} & \text{  1.04} & \text{ 1.028} & \text{ 1.205} & \text{ 1.367} &
   \text{ 1.451} \\
 \text{  142} & \text{ 0.445} & \text{ 0.6053} & \text{ 0.7028} & \text{ 0.755} & \text{ 0.7507} & \text{ 0.8816} & \text{ 1.004}
   & \text{ 1.054} \\
 \text{  150} & \text{ 0.2411} & \text{ 0.387} & \text{ 0.4769} & \text{ 0.539} & \text{ 0.5401} & \text{ 0.6727} & \text{ 0.788}
   & \text{ 0.8107} \\
\hline
 \text{C(r1,r2)} & \text{   94} & \text{  102} & \text{  110} & \text{  118} & \text{  126} & \text{  134} & \text{  142} & \text{
    150} \\
\hline
 \text{   94} & \text{ 5.405} & \text{} & \text{} & \text{} & \text{} & \text{} & \text{} & \text{} \\
 \text{  102} & \text{ 4.453} & \text{ 4.527} & \text{} & \text{} & \text{} & \text{} & \text{} & \text{} \\
 \text{  110} & \text{ 3.355} & \text{ 3.634} & \text{ 3.697} & \text{} & \text{} & \text{} & \text{} & \text{} \\
 \text{  118} & \text{ 2.456} & \text{ 2.735} & \text{ 2.996} & \text{ 3.145} & \text{} & \text{} & \text{} & \text{} \\
 \text{  126} & \text{ 1.933} & \text{ 2.151} & \text{ 2.464} & \text{ 2.743} & \text{ 3.013} & \text{} & \text{} & \text{} \\
 \text{  134} & \text{ 1.505} & \text{ 1.675} & \text{ 1.965} & \text{ 2.257} & \text{ 2.625} & \text{ 2.874} & \text{} & \text{}
   \\
 \text{  142} & \text{ 1.142} & \text{ 1.267} & \text{ 1.511} & \text{ 1.745} & \text{ 2.081} & \text{ 2.435} & \text{ 2.595} &
   \text{} \\
 \text{  150} & \text{ 0.9066} & \text{ 0.9725} & \text{ 1.175} & \text{ 1.364} & \text{ 1.655} & \text{ 1.987} & \text{ 2.224} &
   \text{ 2.351} \\
\hline
\hline
\end{tabular}

\begin{tabular}{ccccccccc}
\hline
\hline
 \text{C(r1,r2)} & \text{   30} & \text{   38} & \text{   46} & \text{   54} & \text{   62} & \text{   70} & \text{   78} & \text{ 
    86} \\
\hline
 \text{   30} & \text{ 88.16} & \text{} & \text{} & \text{} & \text{} & \text{} & \text{} & \text{} \\
 \text{   38} & \text{ 64.43} & \text{ 60.73} & \text{} & \text{} & \text{} & \text{} & \text{} & \text{} \\
 \text{   46} & \text{  46.3} & \text{ 47.45} & \text{ 48.32} & \text{} & \text{} & \text{} & \text{} & \text{} \\
 \text{   54} & \text{ 31.59} & \text{ 34.57} & \text{ 37.63} & \text{ 37.47} & \text{} & \text{} & \text{} & \text{} \\
 \text{   62} & \text{ 22.36} & \text{ 24.87} & \text{  27.6} & \text{ 28.97} & \text{ 27.87} & \text{} & \text{} & \text{} \\
 \text{   70} & \text{ 16.61} & \text{ 18.52} & \text{ 20.92} & \text{ 22.38} & \text{ 22.84} & \text{ 23.49} & \text{} & \text{}
   \\
 \text{   78} & \text{ 12.62} & \text{   14.} & \text{ 15.86} & \text{ 17.49} & \text{ 18.24} & \text{ 19.44} & \text{ 19.94} &
   \text{} \\
 \text{   86} & \text{ 10.01} & \text{ 11.11} & \text{ 12.17} & \text{ 13.61} & \text{ 14.06} & \text{ 15.13} & \text{ 16.58} &
   \text{ 17.06} \\
 \text{   94} & \text{ 7.674} & \text{ 8.367} & \text{ 8.902} & \text{ 10.17} & \text{ 10.58} & \text{ 11.59} & \text{ 12.97} &
   \text{ 14.26} \\
 \text{  102} & \text{ 5.459} & \text{ 5.619} & \text{ 5.887} & \text{  7.01} & \text{ 7.602} & \text{ 8.493} & \text{ 9.906} &
   \text{ 11.18} \\
 \text{  110} & \text{ 4.166} & \text{  3.76} & \text{ 3.783} & \text{ 4.482} & \text{ 5.065} & \text{ 5.691} & \text{ 6.794} &
   \text{ 7.997} \\
 \text{  118} & \text{ 3.224} & \text{ 3.056} & \text{  3.02} & \text{ 3.224} & \text{ 3.587} & \text{ 3.975} & \text{  4.75} &
   \text{ 5.717} \\
 \text{  126} & \text{ 2.576} & \text{ 2.477} & \text{ 2.275} & \text{ 2.214} & \text{ 2.578} & \text{  2.94} & \text{ 3.514} &
   \text{ 4.247} \\
 \text{  134} & \text{  1.27} & \text{ 1.109} & \text{ 1.153} & \text{ 1.215} & \text{ 1.705} & \text{ 2.064} & \text{ 2.476} &
   \text{ 3.051} \\
 \text{  142} & \text{ 0.004535} & \text{ 0.03784} & \text{ 0.3373} & \text{ 0.5719} & \text{ 1.084} & \text{ 1.364} & \text{
   1.764} & \text{ 2.213} \\
 \text{  150} & -0.9506 & -0.5445 & -0.04525 & \text{ 0.1988} & \text{ 0.6778} & \text{  1.04} & \text{ 1.389} & \text{ 1.789} \\
\hline
 \text{C(r1,r2)} & \text{   94} & \text{  102} & \text{  110} & \text{  118} & \text{  126} & \text{  134} & \text{  142} & \text{
    150} \\
\hline
 \text{   94} & \text{ 14.88} & \text{} & \text{} & \text{} & \text{} & \text{} & \text{} & \text{} \\
 \text{  102} & \text{ 12.37} & \text{  12.9} & \text{} & \text{} & \text{} & \text{} & \text{} & \text{} \\
 \text{  110} & \text{ 9.248} & \text{ 10.35} & \text{ 10.57} & \text{} & \text{} & \text{} & \text{} & \text{} \\
 \text{  118} & \text{ 6.905} & \text{  8.08} & \text{ 8.722} & \text{ 9.293} & \text{} & \text{} & \text{} & \text{} \\
 \text{  126} & \text{ 5.098} & \text{ 6.116} & \text{ 6.719} & \text{ 7.655} & \text{ 8.256} & \text{} & \text{} & \text{} \\
 \text{  134} & \text{ 3.662} & \text{ 4.599} & \text{ 5.144} & \text{  6.05} & \text{ 6.983} & \text{ 7.645} & \text{} & \text{}
   \\
 \text{  142} & \text{ 2.589} & \text{ 3.244} & \text{ 3.643} & \text{ 4.449} & \text{ 5.385} & \text{ 6.302} & \text{ 6.868} &
   \text{} \\
 \text{  150} & \text{ 2.005} & \text{ 2.394} & \text{ 2.605} & \text{ 3.227} & \text{  4.12} & \text{ 4.917} & \text{ 5.789} &
   \text{  6.43} \\
\hline
\hline
\end{tabular}
\label{tablecovxinorth}
\caption{Covariance matrices of the spherically averaged correlation function $\xi(s)$ for the LOWZ DR10 NGC (top) and SGC (bottom), derived from one thousand mock galagxy catalogues. The first column and the lines with integers indicate the center of the bins in h$^{-1}$ Mpc. Since the covariance matrix is symmetric only the lower half is displayed, and its values, for clarity, multiplied by $10^6$.}
\end{table*}

\end{document}